\documentclass[%
 reprint,
superscriptaddress,
 amsmath,amssymb,
 aps, prl,
]{revtex4-2}
\usepackage{amsmath,amssymb}
\usepackage{graphicx}
\usepackage{dcolumn}
\usepackage{bm}
\usepackage{multirow}
\usepackage{float}
\usepackage{hhline}
\usepackage{color}
\usepackage{afterpage}
\usepackage{tabularx}
\usepackage{comment}
\usepackage{float}
\usepackage{mathtools}
\usepackage[utf8]{inputenc}
\usepackage[T1]{fontenc}
\usepackage{mathptmx}
\usepackage[normalem]{ulem}
\usepackage{physics}
\usepackage[caption=false]{subfig}

\newcolumntype{C}[1]{>{\hsize=#1\hsize\centering\arraybackslash}X}%
\newcolumntype{L}[1]{>{\hsize=#1\hsize\raggedright\arraybackslash}X}%

\graphicspath{{Figures/}}

\begin{document}

\preprint{AIP/123-QED}

\title{Anomalous Giant Superradiance in Molecular Aggregates Coupled to Polaritons}

\author{Yi-Ting Chuang}
\affiliation{Institute of Atomic and Molecular Sciences, Academia Sinica, Taipei 10617, Taiwan}
\affiliation{Department of Chemistry, National Taiwan University, Taipei 10617, Taiwan}
\author{Liang-Yan Hsu}
\email{lyhsu@gate.sinica.edu.tw}
\affiliation{Institute of Atomic and Molecular Sciences, Academia Sinica, Taipei 10617, Taiwan}
\affiliation{Department of Chemistry, National Taiwan University, Taipei 10617, Taiwan}
\affiliation{Physics Division, National Center for Theoretical Sciences, Taipei 10617, Taiwan}

\begin{abstract}
In this study, we unveil an eccentric superradiance phenomenon in molecular aggregates coupled to surface plasmon polaritons. Through the quantization of electromagnetic fields in media, we demonstrate that superradiance can be significantly enhanced by polaritons and its behavior distinguishably surpasses the Dick's $N$ scaling law. To understand the mechanism of this anomalous phenomenon, we derive an analytical expression of the superradiance rate, which is general for molecular aggregates in arbitrary dispersive and absorbing media. Furthermore, we show the importance of intermolecular distance for this extraordinary superradiance.
\end{abstract}

\maketitle

\textit{Introduction}.--- 
In 1954, Dicke in his seminal work introduced the concept of superradiance \cite{Dicke1954}, where a group of closely spaced molecules can emit light coherently, resulting in a collective emission rate surpassing that of a single monomer. Since then, this intriguing phenomenon has been extensively investigated in various systems such as atoms \cite{Gross1976,DeVoe1996,Goban2015}, molecular ensembles \cite{Ulusoy2020,Chen2022}, quantum dots \cite{Scheibner2007,Tighineanu2016}, Bose-Einstein condensates \cite{Joseph2004,Rezende2009,Mi2021}, and particularly molecular aggregates \cite{Kasha1958,Kasha1959,Kasha1963,Kasha1965,Grad1988,Spano1989,Spano2010,Doria2018}. 
\textcolor{black}{Among these} pioneering stduies\textcolor{black}{,} Kasha \textit{\textit{et al}}. \textcolor{black}{conducted a series of works on excitons formed by molecular aggregates based on considering only the Coulombic dipole-dipole interactions between the monomers and successfully explained experimental luminescence spectra} in the 1950s to 1960s \cite{Kasha1958,Kasha1959,Kasha1963,Kasha1965}. \textcolor{black}{Another} significant advancement was made by Spano \textcolor{black}{and Mukamel} in 1989 \cite{Spano1989}. \textcolor{black}{They applied a full quantum electrodynamics (QED) framework to study 
molecular aggregates and derived the collective emission rate of the superradiant state and the corresponding energy level in free space.} 

\begin{figure}[t!]
    \centering
    \includegraphics[width=0.48\textwidth]{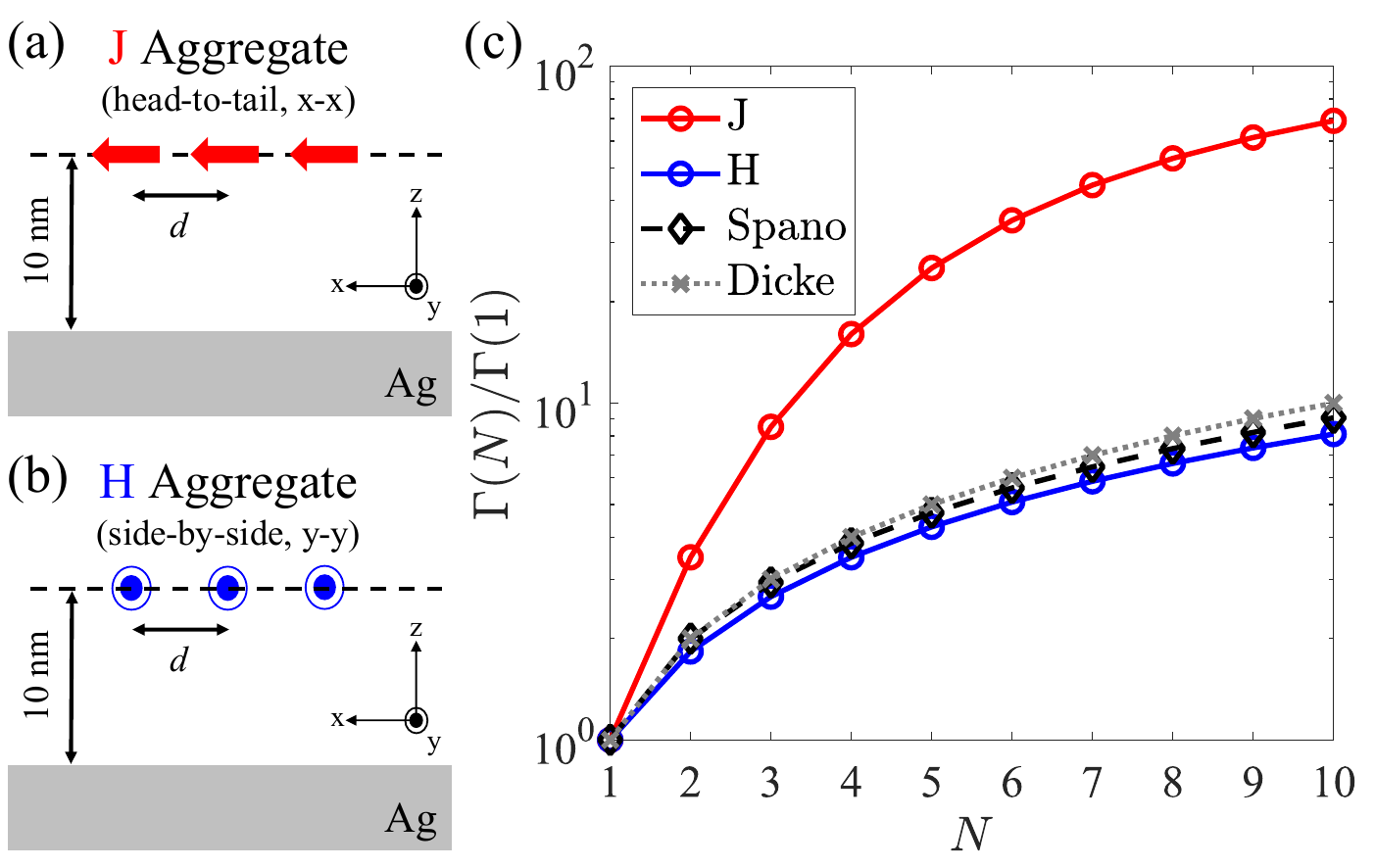} 
    \caption{Schematic illustration of (a) J aggregate and (b) H aggregate above a silver surface. (c) $N$ dependence of superradiance rate enhancement calculated based on Eq.~(\ref{Eq:QuantumDynamics}). The red and blue lines with circles correspond to the results of J and H aggregates, respectively. The black dashed line with diamonds corresponds to the enhancement in free space studied by Spano \textit{et al}. \cite{Spano1989}, i.e., $\Gamma(N)/\Gamma(1) = \abs{\sum_{\alpha} U_{\alpha 1}}^2$, and the gray dotted line with crosses corresponds to the $N$-fold enhancement proposed by Dicke \cite{Dicke1954}.}
    \label{Fig:Illustration}
\end{figure}

\textcolor{black}{Vacuum fluctuations due to confined electromagnetic fields, such as polaritons and cavity photons, are considered as a novel approach to alter molecular properties and have been employed to modify resonance energy transfer \cite{Andrew2004,Coles2014,Zhong2017,Wu2018} and chemical reactions \cite{Hutchison2012,Semenov2019,Thomas2019}.}  
\textcolor{black}{In fact, superradiance led by cavity photons has been extensively investigated, but the impact of polaritons on superradiance in molecular aggregates has received less attention due to the difficulty to appropriately describe the quantization of electromagnetic fields in media.}
\textcolor{black}{To further unveil how polaritons influence superradiance,} in this letter, we \textcolor{black}{consider polaritons formed of photons and dielectric media and} explore superradiance in molecular aggregates coupled to surface plasmon polaritons (SPPs), i.e., molecular aggregates positioned above a plasmonic surface \textcolor{black}{as shown in Figs.~\ref{Fig:Illustration}(a) and \ref{Fig:Illustration}(b)}, based on the equation of motion \cite{Chuang2024} derived from macroscopic QED (MQED) \cite{Gruner1996,Dung1998,Buhmann2012}. We believe that MQED serves as an appropriate theoretical framework for studying superradiance in dielectric environments because (i) MQED extends molecular QED into regimes including inhomogeneous, dispersive, and absorbing media \cite{Gruner1996}, (ii) MQED is a full \textit{ab initio} method without free parameters \cite{Herrera2020}, and (iii) \textcolor{black}{several experimental observations are in good agreement with the theories derived from MQED} \cite{Zhang2014,Wang2019,Boddeti2022}. \textcolor{black}{Based on the MQED approach, this study not only expands the theory of superradiance in molecular aggregates from free space to complex dielectric environments but also puts forward a new way for manipulating molecular superradiance.}

\textit{Model Hamiltonian}.---
We consider a molecular aggregate formed of $N$ identical monomers coupled to polaritons (photons dressed by the dielectric environment) at low temperatures, where each monomer is treated as an electronically two-level system. Since we focus on the effect of the dielectric environment on superradiance in this work, we do not consider the influence of phonon interactions (negligible at low temperatures \cite{Spano1989}), charge transfer (negligible at sufficient intermolecular distance \cite{Hestand2017}), and disorder. According to MQED in the multipolar coupling framework under the electric-dipole approximation, the total Hamiltonian $\hat{H}$ can be expressed as $\hat{H} = \hat{H}_\mathrm{M} + \hat{H}_\mathrm{P} + \hat{H}_\mathrm{I}$. The molecular Hamiltonian for \textcolor{black}{$N$ identical monomers is} $\hat{H}_\mathrm{M} = \sum^{N}_{\alpha=1} \hbar\omega_{\textcolor{black}{\rm{M}}} \hat{\sigma}^{+}_\alpha \hat{\sigma}^{-}_\alpha$, where $\omega_{\textcolor{black}{\rm{M}}}$ is the transition frequency of all the monomers, and $\hat{\sigma}^{+}_\alpha$ ($\hat{\sigma}^{-}_\alpha$) is the raising (lowering) operator of $\alpha$. The polaritonic Hamiltonian is \textcolor{black}{described by} $\hat{H}_\mathrm{P} = \int \mathrm{d}\mathbf{r} \int_{0}^{\infty} \mathrm{d}\omega \, \hbar\omega \,\mathbf{\hat{f}}^\dagger(\bf{r},\omega)\cdot\mathbf{\hat{f}}(\bf{r},\omega)$, where $\mathbf{\hat{f}}^\dagger(\mathbf{r},\omega)$ and $\mathbf{\hat{f}}(\mathbf{r},\omega)$ are the creation and annihilation operators of \textcolor{black}{the bosonic vector fields}. 
\textcolor{black}{Note that we include the counter-rotating interactions into the interaction Hamiltonian $\hat{H}_\mathrm{I} = - \sum_{\alpha} \hat{\boldsymbol{\mu}}_{\alpha} \cdot \hat{\mathbf{F}}(\mathbf{r}_\alpha)$ because of the counter-rotating interactions as an essential component in the intermolecular dipole-dipole interaction \cite{Chuang2024}.}
The transition dipole operator $\hat{\boldsymbol{\mu}}_\alpha$ can be expressed as $\hat{\boldsymbol{\mu}}_\alpha = \boldsymbol{\mu}_\alpha [\hat{\sigma}^{+}_{\alpha} + \hat{\sigma}^{-}_{\alpha}]$, where $\boldsymbol{\mu}_\alpha$ represents the transition dipole moment of $\alpha$. The field operator $\hat{\mathbf{F}}(\mathbf{r}_\alpha)$ can be expressed as $\hat{\mathbf{F}}(\mathbf{r}_\alpha) = \int \dd{\mathbf{r}} \int_0^\infty \dd{\omega}  \overline{\overline{\mathcal{G}}}(\mathbf{r}_\alpha,\mathbf{r},\omega) \cdot \hat{\mathbf{f}}(\mathbf{r},\omega) + \mathrm{H.c.}$, where $\overline{\overline{\mathcal{G}}}(\mathbf{r}_\alpha,\mathbf{r},\omega) = i\sqrt{\frac{\hbar}{\pi\varepsilon_0}} \frac{\omega^2}{c^2} \sqrt{\mathrm{Im} \left[ \varepsilon_\mathrm{r}(\mathbf{r},\omega) \right]} \, \overline{\overline{\mathbf{G}}}(\mathbf{r}_\alpha,\mathbf{r},\omega)$. $\varepsilon_0$, $\varepsilon_\mathrm{r}(\mathbf{r},\omega)$, and $c$ are the permittivity of free space, the relative permittivity, and the speed of light in free space, respectively; $\overline{\overline{\mathcal{G}}}(\mathbf{r}_\alpha,\mathbf{r},\omega)$ is an auxiliary tensor defined in terms of the dyadic Green's function $\overline{\overline{\mathbf{G}}}(\mathbf{r},\mathbf{r'},\omega)$ that satisfies macroscopic Maxwell’s equations $\left[ \frac{\omega^2}{c^2}\varepsilon_\mathrm{r}(\mathbf{r},\omega) - \nabla \times \nabla \times \right] \overline{\overline{\mathbf{G}}}(\mathbf{r},\mathbf{r'},\omega) = -\mathbf{\overline{\overline{I}}}_3 \delta(\mathbf{r}-\mathbf{r'})$ with $\mathbf{\overline{\overline{I}}}_3$ being the three-dimensional identity matrix.

\textit{State vector and quantum dynamics}.---
To adequately incorporate the effect of counter-rotating interactions, we adopt an extended Wigner-Weisskopf wave function ansatz to represent the state vector $\ket{\Psi(t)}$ of the total system \textcolor{black}{(see the details in Supplemental Material
(SM) \cite{SM})}. The state vector includes the following states: (i) a single excited molecule with zero polariton, i.e., $\ket{\mathrm{E_{\alpha}}} \ket{\left\{ 0 \right\}} = \hat{\sigma}^{+}_\alpha \ket{\mathrm{G}} \ket{\left\{ 0 \right\}}$, (ii) zero excited molecule with one polariton, i.e., $\ket{\mathrm{G}} \ket{\left\{1_l(\mathrm{\bf{r}}, \omega)\right\}} = \hat{f}^\dagger_l\left( \mathbf{r}, \omega \right) \ket{\mathrm{G}} \ket{\{0\}}$, and (iii) two excited molecules with one polariton, i.e., $\ket{\mathrm{E_{\alpha\beta}}} \ket{\left\{1_l(\mathrm{\bf{r}}, \omega) \right\}} = \hat{\sigma}^{+}_\alpha \hat{\sigma}^{+}_{\beta \neq \alpha} \hat{f}^\dagger_l\left( \mathbf{r}, \omega \right) \ket{\mathrm{G}} \ket{\{0\}}$. Here, $\ket{\mathrm{G}}$ denotes that all the molecules are in their electronically ground states, and $\lvert\{0\}\rangle$ denotes the zero-polariton state. The probability amplitudes of $\lvert\Psi(t)\rangle$ for $\ket{\mathrm{E}_\alpha} \lvert\{0\}\rangle$, $\ket{\mathrm{G}} \lvert\{1_l(\mathbf{r},\omega)\}\rangle$ and $\ket{\mathrm{E}_{\alpha\beta}} \lvert\{1_l(\mathbf{r},\omega)\}\rangle$ are denoted as $C^\mathrm{E_\alpha,\left\{0\right\}}(t)$, $C^{\mathrm{G},\{1_l\}}(\mathbf{r},\omega,t)$ and $C^{\mathrm{E}_{\alpha\beta},\{1_l\}}(\mathbf{r},\omega,t)$, respectively. 

Substituting the state vector into the Schr\"{o}dinger equation, i.e., $i\hbar \partial \ket{\Psi(t)}/ \partial t = \hat{H} \ket{\Psi(t)}$, with the initial condition that there are no polaritons at $t=0$, i.e., $C^{\mathrm{G},\left\{1_l\right\}}(\mathbf{r},\omega,t=0) = C^{\mathrm{E_{\alpha \beta}}, \left\{1_l\right\}}(\mathbf{r},\omega,t=0) = 0$, we obtain the equation of motion of the excited-state molecules as \cite{Chuang2024}
\begin{align}
\nonumber
    & \dv{t} C^{\mathrm{E_{\alpha}},\left\{0\right\}}(t) = \\
\nonumber
    & - \sum_{\beta} \int_0^t \dd{t'} \int_0^\infty \dd{\omega} J_{\alpha \beta}(\omega) e^{-i \left( \omega - \omega_0 \right) (t-t')} C^{\mathrm{E_{\beta}},\left\{0\right\}}(t') \\
\nonumber
    & - \sum_{\beta\neq\alpha} \int_0^t \dd{t'} \int_0^\infty \dd{\omega} J_{\beta \beta}(\omega) e^{-i \left( \omega + \omega_0 \right) (t-t')}  C^{\mathrm{E_{\alpha}},\left\{0\right\}}(t') \\
    & - \sum_{\beta\neq\alpha} \int_0^t \dd{t'} \int_0^\infty \dd{\omega} J_{\beta \alpha}(\omega) e^{-i \left( \omega + \omega_0 \right) (t-t')}  C^{\mathrm{E_{\beta}},\left\{0\right\}}(t'),
\label{Eq:QuantumDynamics}
\end{align}
where $\overline{\overline{\mathbf{J}}}(\omega)$ is the generalized spectral density \cite{Chuang2022} whose element $J_{\alpha \beta}(\omega)$ is defined as
\begin{align}
    J_{\alpha \beta}(\omega) & = \frac{\omega^2}{\pi \hbar \epsilon_0 c^2} \boldsymbol{\mu}_\alpha \cdot \mathrm{Im} \overline{\overline{\mathbf{G}}}(\mathbf{r}_\alpha, \mathbf{r}_\beta,\omega) \cdot \boldsymbol{\mu}_\beta.
\label{Eq:SpectralDensity}
\end{align}
More details about the derivation of the EOM can be found in SM \cite{SM}. Eq.~(\ref{Eq:QuantumDynamics}) is a set of coupled integro-differential equations, and it can capture both Markovian and non-Markovian dynamics of molecules coupled to polaritons from weak to strong coupling regimes. Using the conventional Markov approximation, Eq.~(\ref{Eq:QuantumDynamics}) can be reduced to a set of coupled differential equations, which is equivalent to the Lindblad-type master equation \cite{Dung2002_2,Masson2020} \textcolor{black}{in the single-excitation subspace}, and their equivalence is demonstrated in SM \cite{SM}. \textcolor{black}{How to solve the dyadic Green's functions in Eq.~(\ref{Eq:SpectralDensity}) for molecules above a silver surface can be found in our previous work \cite{Wu2018}. In addition, the dielectric function of silver we used is based on an analytical model \cite{Melikyan2014} that accurately fits the experimental data \cite{Johnson1972} and ensures the Kramers-Kronig relation (causality). }

\begin{figure*}[!t]
    \centering
    \includegraphics[width=1\textwidth]{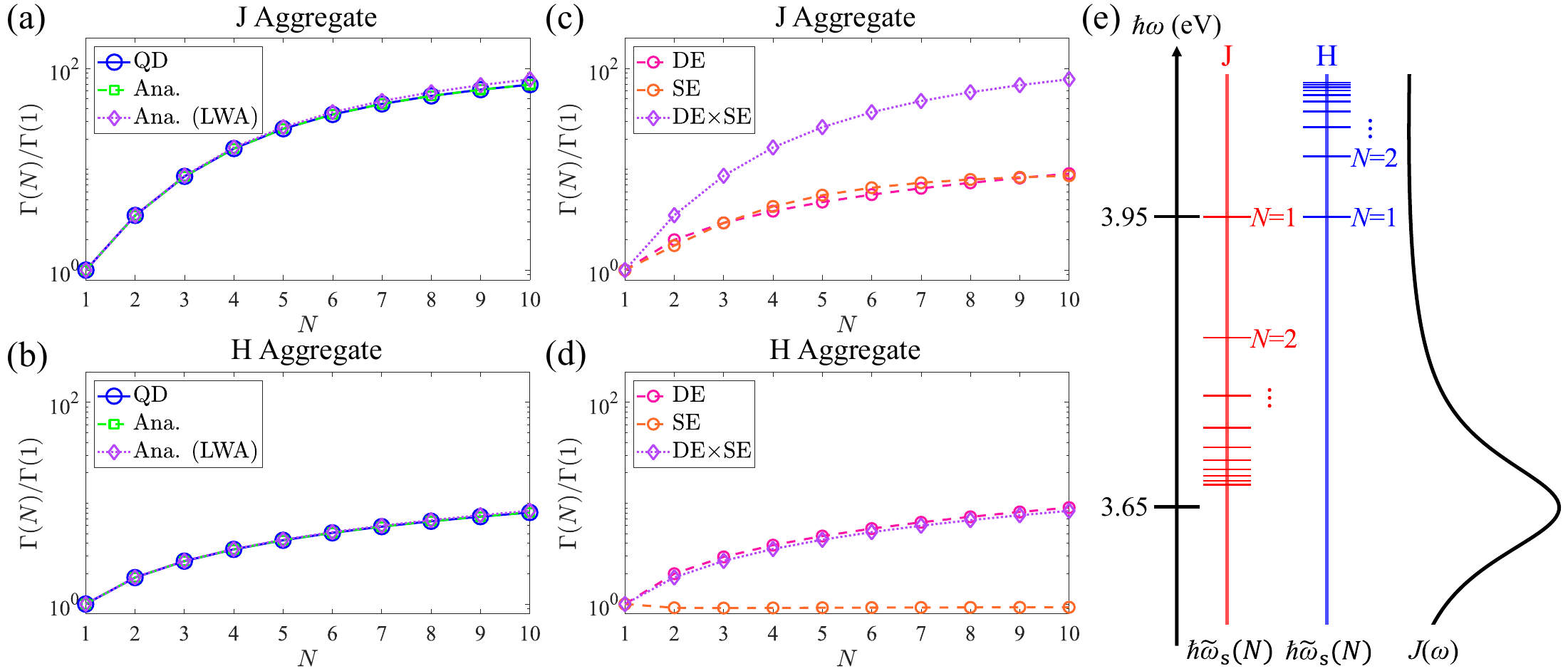} 
    \caption{(a), (b) $N$ dependence of superradiance rate enhancement in (a) J aggregates and (b) H aggregates calculated from the quantum dynamical (QD) simulations based on Eq.~(\ref{Eq:QuantumDynamics}), the analytical expression of the superradiance rate (Ana.) in Eq.~(\ref{Eq:GeneralRate}), and the analytical expression of the superradiance rate with the long-wavelength approximation (Ana. (LWA)) in Eq.~(\ref{Eq:GeneralRate_LWA}). (c), (d) Decomposition of the rate enhancement into the dipole enhancement (DE) and the spectral enhancement (SE) based on Eqs.~(\ref{Eq:Enhancement})-(\ref{Eq:SpectralEnhancement}) for (c) J aggregates and (d) H aggregates. Note that DE is exactly equivalent to the enhancement in free space. (e) The profile of the spectral density $J(\omega)$, as indicated by the black line, and the energy levels of the superradiant states $\hbar \tilde{\omega}_{\textcolor{black}{\rm{s}}}(N)$, as indicated by the red and blue lines for J and H aggregates, respectively.}
    \label{Fig:Analytical}
\end{figure*}

\textit{Superradiance in one-dimensional linear aggregates}.---
In this work, we concentrate on one-dimensional linear aggregates weakly coupled to surface plasmon polaritons, i.e., above a silver surface, with equal spacing between two adjacent monomers. We explore two distinct configurations of linear aggregates: (i) J aggregates, characterized by the head-to-tail alignment of monomers' transition dipole moments, as depicted in Fig.~\ref{Fig:Illustration}(a), and (ii) H aggregates, where monomers' transition dipole moments align side-by-side, as illustrated in Fig.~\ref{Fig:Illustration}(b). Since we are interested in examining the influence of polaritons (the dielectric environment) on the superradiance rate, we initiate our system in the superradiant (bright) state of the molecular aggregate in free space \cite{Spano1989} and perform numerical simulations based on the equation of motion presented in Eq.~(\ref{Eq:QuantumDynamics}). To determine the coefficients of the superradiant state, we follow the same procedure as done by Spano \textit{et al} \cite{Spano1989} and identify the unitary matrix $\overline{\overline{\mathbf{U}}}$ that diagonalizes the free-space dipole-dipole interaction matrix  $\overline{\overline{\mathbf{V}}}\vphantom{a}^0$, where the matrix element $V^0_{\alpha \beta} = \frac{-\omega_0^2}{\varepsilon_0 c^2}\boldsymbol{\mu}_{\alpha} \cdot \mathrm{Re}\overline{\overline{\mathbf{G}}}\vphantom{a}^0(\mathbf{r}_\alpha,\mathbf{r}_\beta,\omega_{\textcolor{black}{\rm{M}}}) \cdot \boldsymbol{\mu}_{\beta}$ for $\alpha\neq\beta$ and $V^0_{\alpha \beta} = 0$ for $\alpha=\beta$, with $\overline{\overline{\mathbf{G}}}\vphantom{a}^0(\mathbf{r}_\alpha,\mathbf{r}_{\beta\neq\alpha},\omega_{\textcolor{black}{\rm{M}}})$ being the two-point free-space dyadic Green's functions \cite{Chuang2024}. 
\textcolor{black}{After the diagonalization, the superradiant state $\ket{\tilde{\mathrm{E}}_{\rm{s}}}$ can be expressed as a linear combination of the monomer excited states $\ket{\mathrm{E}_{\alpha}}$,
\begin{align}
 \ket{\tilde{\mathrm{E}}_{\rm{s}}} = \sum_{\alpha} U_{\alpha \rm{s}} \ket{\mathrm{E}_{\alpha}},
\label{Eq:CoefficientAggreteState}
\end{align}
with the transition energy
\begin{align}
    \hbar\tilde{\omega}_{\rm{s}} & = \hbar \omega_{\textcolor{black}{\rm{M}}} + \hbar\tilde{\Delta}_{\rm{s}},
\label{Eq:EnergyAggreteState}
\end{align}
where $\hbar\tilde{\Delta}_{\rm{s}}  = \sum_{\alpha,\beta} U_{\alpha \rm{s}}^* U_{\beta \rm{s}} V^0_{\alpha \beta}$.
The superradiant state is identified by a uniform phase across all components in Eq.~(\ref{Eq:CoefficientAggreteState}), i.e., $U_{\alpha \mathrm{s}}$ with the same sign for all $\alpha$.}
Numerically, we determine the effective decay rate of the superradiant state by calculating Eq.~(\ref{Eq:QuantumDynamics}) and fitting $\ln{\left[ P_\mathrm{tot}(t) \right]} = -\Gamma_\mathrm{fit}t$, where $P_\mathrm{tot}(t) = \sum_\alpha \abs{C^{\mathrm{E_{\alpha}},\left\{0\right\}}(t)}^2$ represents the total excited-state population. The coefficient of determination, $R^2$, for all fits in this study exceeds 0.99, confirming that a simple exponential decay model captures the dynamics of the superradiant state. 

In Fig.~\ref{Fig:Illustration}(c), we illustrate the significant increase in the superradiance rate, quantified as the decay rate of $N$ monomers, $\Gamma(N)$, normalized by the decay rate of a single monomer, $\Gamma(1)$. We set the molecule-surface distance at $h = 10$ nm (within the weak-coupling regime \cite{Wang2020}), and the intermolecular distance at $d = 1$ nm, with the transition dipole moment magnitude fixed at $\mu = 10$ Debye. \textcolor{black}{We chose the off-resonant condition between the monomer and the SPP mode ($\hbar\omega_{\rm{M}}=3.95$ eV and $\hbar\omega_\mathrm{SPP}=3.65$ eV) because we found no evident superradiance in the resonant condition $\hbar\omega_{\rm{M}}=\hbar\omega_\mathrm{SPP}$.} We compare our findings with the superradiance rate enhancement of aggregates in free space as studied by Spano \textit{\textit{et al}}. \cite{Spano1989}, which can be expressed as $\Gamma(N)/\Gamma(1) = \abs{\sum_{\alpha} U_{\alpha \mathrm{s}}}^2$, and the classical $N$-fold increase in the single-photon superradiance rate proposed by Dicke, i.e., $\Gamma(N)/\Gamma(1) = N$. Our results reveal that J aggregates (red line with circles) coupled to SPPs exhibit a rate enhancement that significantly exceeds both the free-space prediction by Spano \textit{et al}. and Dicke's $N$-fold increase. Specifically, for $N=10$, the superradiance rate is about 70 times that of a single monomer, marking an unprecedented observation in molecular aggregates. Moreover, even for a dimer, $N=2$, the enhancement surpasses threefold, differentiating it from the conventional expectation of a two-times increase. Conversely, when considering H aggregates coupled to SPPs (blue line with circles), the surge in the superradiance rate is not observed. Instead, the rate enhancement is marginally lower than that in free space.

\textit{General formula of superradiance rate and physical interpretation}.---
To delve into the underlying physics of \textcolor{black}{the giant superradiance as depicted by the red line in Fig.~\ref{Fig:Illustration}(c)}, we have formulated an analytical expression for the superradiance rate $\Gamma_\mathrm{sup}$, which can be expressed as [see SM \cite{SM} for the derivation]
\begin{align}
    \Gamma_\mathrm{sup} = 2 \pi \sum_{\alpha, \beta} U^*_{\alpha \mathrm{s}} U_{\beta \mathrm{s}} J_{\alpha \beta}(\tilde{\omega}_\mathrm{s}).
\label{Eq:GeneralRate}
\end{align}
Recall that $U_{\alpha \mathrm{s}}$ and $\tilde{\omega}_\mathrm{s}$ represent the coefficient and the transition frequency of the superradiant state in Eqs.~(\ref{Eq:CoefficientAggreteState}) and (\ref{Eq:EnergyAggreteState}), respectively. Note that Eq.~(\ref{Eq:GeneralRate}) is a general formula that applies to any arbitrary dielectric environment and is not confined solely to linear aggregates.

\textcolor{black}{Moreover, if we consider that the monomers are sufficiently close, we can further make two approximations to Eq.~(\ref{Eq:GeneralRate}). First, all monomers are subject to a uniform electric field led by the dielectric environment, i.e., $J_{\alpha \alpha}(\omega) \approx J_{\beta \beta}(\omega)$ for any monomers $\alpha$ and $\beta$. Note that $J_{\alpha \alpha}(\omega) = J_{\beta \beta}(\omega)$ in the present study due to the translation symmetry of each monomer above a surface. Second, the off-diagonal elements of the generalized spectral density are similar to the diagonal elements, i.e., $J_{\alpha \beta}(\omega) \approx J_{\alpha \alpha}(\omega)$. As a result, we can denote $J(\omega) = J_{\alpha \alpha}(\omega) = J_{\alpha \beta}(\omega)$ and reduce Eq.~(\ref{Eq:GeneralRate}) to
\begin{align}
    \Gamma_\mathrm{sup} \approx \abs{\sum_{\alpha} U_{\alpha \mathrm{s}}}^2 \times 2 \pi J(\tilde{\omega}_\mathrm{s}),
\label{Eq:GeneralRate_LWA}
\end{align}
and we referred to Eq.~(\ref{Eq:GeneralRate_LWA}) as the analytical solution with the long-wavelength approximation (LWA).}

In Figs.~\ref{Fig:Analytical}(a) and \ref{Fig:Analytical}(b), we \textcolor{black}{compare the superradiance rate enhancement according to three different methods:} (i) the numerical \textcolor{black}{quantum dynamical simulations based on Eq.~(\ref{Eq:QuantumDynamics})}, as denoted by QD, (ii) \textcolor{black}{the analytical solution based on Eq.~(\ref{Eq:GeneralRate})}, as denoted by Ana., and (iii) the analytical solution with the LWA based on Eq.~(\ref{Eq:GeneralRate_LWA}), as denoted by Ana. (LWA). For both J aggregates in Fig.~\ref{Fig:Analytical}(a) and H aggregates in Fig.~\ref{Fig:Analytical}(b), 
\textcolor{black}{the good agreement between QD and Ana. (not only the rate enhancements but also the rates themselves) shows the reliability of our general formula of rate in Eq.~(\ref{Eq:GeneralRate}).}
\textcolor{black}{In addition, the slight deviation between QD and Ana. (LWA) indicates that the long-wave approximation successfully captures the emission rates of the superradiant states.} Note that the breakdown of the LWA occurs at a large intermolecular distance, in which $J_{\alpha \beta}(\omega) \not\approx J_{\alpha\alpha}(\omega)$. Since $\sqrt{J_{\alpha\alpha}(\omega) J_{\beta \beta}(\omega)} \geq J_{\alpha \beta}(\omega)$ \cite{Canaguier-Durand2019}, the LWA tends to overestimate the rate enhancement for large molecular aggregates. 

\begin{figure}[!t]
    \centering
    \includegraphics[width=0.47\textwidth]{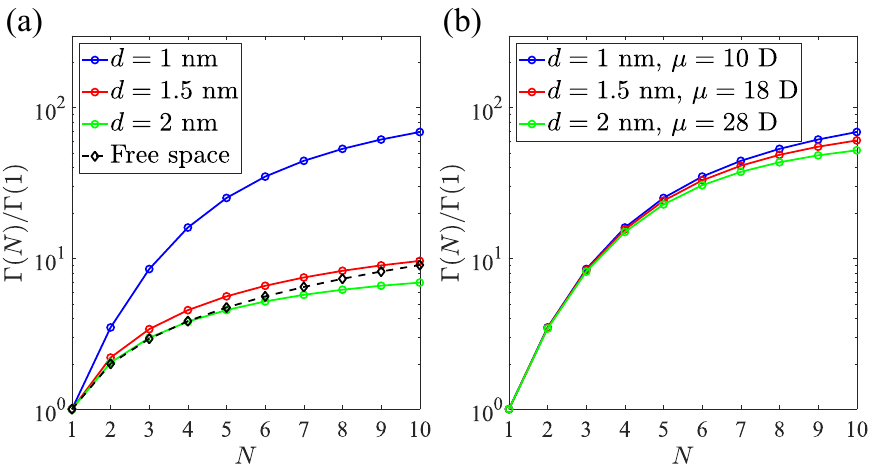} 
    \caption{(a) Intermolecular-distance dependence of the rate enhancement in J aggregates coupled to SPPs calculated based on Eq.~(\ref{Eq:QuantumDynamics}) and in free space. The enhancements in free space for different intermolecular distances (1 nm, 1.5 nm, 2 nm) are almost the same, so we only show one black dashed line with diamonds. (b) Rate enhancement in J aggregates with a fixed $\mu^2 / d^3$.}
    \label{Fig:Distance}
\end{figure}

Based on Eq.~(\ref{Eq:GeneralRate_LWA}), we can further express the enhancement of the superradiance rate as
\begin{align}
    \frac{\Gamma(N)}{\Gamma(1)} & \approx \mathrm{DE} \times \mathrm{SE},
\label{Eq:Enhancement} \\
    \mathrm{DE} &= \abs{\sum_{\alpha} U_{\alpha \mathrm{s}}}^2,
\label{Eq:DipoleEnhancement} \\
    \mathrm{SE} &= \frac{J(\tilde{\omega}_\mathrm{s})}{J(\omega_\mathrm{M})},
\label{Eq:SpectralEnhancement}
\end{align}
where DE refers to the “\textit{dipole enhancement}” and SE refers to the “\textit{spectral enhancement}”. The factor DE describes the increase in the square of the effective transition dipole moment associated with the superradiant state compared to that of a single monomer, and it is equivalent to the superradiance rate enhancement in the free-space condition discussed by Spano \textit{et al} \cite{Spano1989}.
The factor SE reflects that the enhancement of the superradiance rate is associated with the change of the spectral density (photonic density of states) due to the energy shift in the aggregate's superradiant state. \textcolor{black}{Note that SE can be neglected (SE $\approx 1$) when molecules are in free space (flat spectral density) or the intermolecular interactions are weak ($\tilde{\omega}_\mathrm{s} \approx \omega_\mathrm{M}$).} 

In a dielectric environment, SE plays a key role in the superradiance rate enhancement because the spectral density can vary enormously with frequency. In Figs.~\ref{Fig:Analytical}(c) and \ref{Fig:Analytical}(d), we plot DE and SE (correspond to the cases in Figs.~\ref{Fig:Analytical}(a) and \ref{Fig:Analytical}(b), respectively) separately to demonstrate their effects on the enhancement of the superradiance rate. For both J and H aggregates, DEs are almost the same (the difference is smaller than 0.1\%)\textcolor{black}{, but their SEs can be very different.} For example, 
for J aggregates, SE can increase by an order of magnitude, while for H aggregates, SE remains almost constant (slightly smaller than 1). 

The variation in SE can be understood through the energy levels of the superradiant states and the profile of the spectral density $J(\omega)$, as depicted in Fig.~\ref{Fig:Analytical}(e). For J aggregates, \textcolor{black}{the} increase in the number of monomers \textcolor{black}{shifts} the energy of the superradiant state $\hbar \tilde{\omega}_\mathrm{s}$ to lower energy levels, as indicated by the red vertical line. This downward energy shift of the superradiant state leads to an enlargement of $J(\tilde{\omega}_\mathrm{s})$, i.e., the resonance between the superradiant state and the SPP mode, and hence an increasing SE. On the contrary, for H aggregates, \textcolor{black}{the} increase in the number of monomers \textcolor{black}{shifts} the energy of the superradiant state $\hbar \tilde{\omega}_\mathrm{s}$ to higher energy levels, as indicated by the blue vertical line. This upward energy shift of the superradiant state leads to an reduction of $J(\tilde{\omega}_\mathrm{s})$, i.e., the superradiant state away from the SPP mode, and hence a decreasing SE. \textcolor{black}{Therefore, if we consider the on-resonant condition where the momoner is resonant with the SPP mode, i.e., $\hbar \omega_\mathrm{M} = \hbar \omega_\mathrm{SPP} = 3.65$, we cannot observe the extremely large enhancement of the superradiance rate because the energy levels of the superradiant states in both J and H aggregates shift to positions off-resonant with the SPP mode, leading to a diminished $J(\tilde{\omega}_\mathrm{s})$ as compared to a single monomer and consequently, a decrease in SE.}

The analytical expression of the rate enhancement accurately captures the observed anomalous phenomenon and provides a principle of modifying superradiance rates. To further simplify the calculation of the superradiance rate enhancement, we derive easy-to-use formulae for estimating the magnitude of DE and SE of a linear aggregate near a dielectric surface based on a series of approximations in SM \cite{SM}.

\textit{Effects of intermolecular distance}.---
As shown in Fig.~\ref{Fig:Analytical}, SE is highly contingent on the energy level of the superradiant state, which is associated with the free-space dipole-dipole interactions among the monomers within an aggregate. These dipole-dipole interactions are a function of the intermolecular distance as well as the magnitude of the transition dipole moment. In Fig.~\ref{Fig:Distance}(a), we examine how changes in intermolecular distance affect the rate enhancement in J aggregates. At a separation of 1.5 nm between adjacent monomers, the previously significant superradiance rate enhancement is markedly diminished. Further increasing the intermolecular distance to 2 nm entirely eliminates the remarkable enhancement, and the enhancement even falls below that seen in free space. 

To reinstate the giant enhancement in superradiance rate at a large intermolecular distance, one can increase the transition dipole moment strength of each monomer to compensate for the reduced dipole-dipole interaction. In the non-retarded limit ($\omega_0 r_{\alpha \beta} / c \ll 1$) as considered here, the free space dipole-dipole interaction can be simplified to its electrostatic limit, leading to a relationship
\begin{align}
    \mu^2 / r_{\alpha \beta}^3 \propto V^0_{\alpha \beta}.
\end{align}
Based on this relationship, we maintain a constant value for $\mu^2 / d^3$ across different configurations ($d$ is the distance between two adjacent monomers), as demonstrated in Fig.~\ref{Fig:Distance}(b). It is shown that the notable enhancement in superradiance rate is again restored. In conclusion, we have shown that the giant superradiance can be controlled by tuning the intermolecular distance and the transition dipole moment strength.

\textit{Conclusions}.---
In this study, we presented an anomalous giant superradiance in molecular aggregates coupled to SPPs through the quantization of electromagnetic fields in media.
\textcolor{black}{This giant superradiance  distinguishably deviates the behavior proposed by Spano \textit{et al.} and the Dick's scaling law, and its magnitude can be roughly an order of magnitude larger than the traditional superradiance even when the number of monomers in the molecular aggregate is small, e.g., $N = 10$.} To understand the \textcolor{black}{intrinsic} mechanism, we analytically derived a general formula of superradiance rate in any arbitrary dielectric environment, which not only confirms the observed anomalous phenomenon but also offers \textcolor{black}{the principle of altering superradiance rates.} Furthermore, we showed that the rate enhancement can be approximately decomposed into two factors, including the dipole enhancement and the spectral enhancement. \textcolor{black}{The former covers the classic mechanism proposed by Spano \textit{et al.}}, while the latter is the origin of the giant superradiance caused by SPPs. \textcolor{black}{Our results showed that the spectral enhancement can be controlled by varying the intermolecular distance and transition dipole moment.} \textcolor{black}{We believe that this study opens up new directions for exploring the fundamental mechanisms of superradiance in molecular aggregates and their potential applications.}

\begin{acknowledgments}
We thank Academia Sinica (AS-CDA-111-M02), National Science and Technology Council (111-2113-M-001-027-MY4) and Physics Division, National Center for Theoretical Sciences (112-2124-M-002-003) for the financial support.
\end{acknowledgments}

\bibliographystyle{apsrev4-2}

%

\end{document}